\documentclass[preprint,journal]{vgtc}       

\ieeedoi{10.1109/TVCG.2019.2934660}

\ifpdf
  \pdfoutput=1\relax                   
  \usepackage{graphicx}                
  \DeclareGraphicsExtensions{.pdf,.png,.jpg,.jpeg} 
\else
  \usepackage{graphicx}                
  \DeclareGraphicsExtensions{.eps}     
\fi%

\graphicspath{{figures/}{pictures/}{images/}{./}} 

\usepackage{microtype}                 
\PassOptionsToPackage{warn}{textcomp}  
\usepackage{textcomp}                  
\usepackage{mathptmx}                  
\usepackage{times}                     
\usepackage{cite}                      
\usepackage{tabu}                      
\usepackage{booktabs}                  
\usepackage{xcolor}
\usepackage{paralist}

\newcommand{\tabincell}[2]{\begin{tabular}{@{}#1@{}}#2\end{tabular}} 

\onlineid{1040}

\vgtccategory{Research}
\vgtcpapertype{Application}

\title{sPortfolio: Stratified Visual Analysis of Stock Portfolios}

\author{
Xuanwu Yue, Jiaxin Bai, Qinhan Liu, Yiyang Tang, Abishek Puri, Ke Li, Huamin Qu
}
\authorfooter{

\item
Xuanwu Yue is with the Hong Kong University of Science and Technology and Sinovation Ventures AI Institute. Email: xuanwu.yue@connect.ust.hk

\item
Jiaxin Bai, Qinhan Liu, Yiyang Tang, Abishek Puri and Huamin Qu are with the Hong Kong University of Science and Technology. Email:\{jbai, qliuan, ytangan, apuri, huamin\} @connect.ust.hk 

\item
Ke Li is with the RiceQuant Co. Ltd. Email: lk@ricequant.com

}

\shortauthortitle{Biv \MakeLowercase{\textit{et al.}}: Global Illumination for Fun and Profit}

\abstract{Quantitative Investment, built on the solid foundation of robust financial theories, is at the center stage in investment industry today. The essence of quantitative investment is the multi-factor model, which explains the relationship between the risk and return of equities. However, the multi-factor model generates enormous quantities of factor data, through which even experienced portfolio managers find it difficult to navigate. This has led to portfolio analysis and factor research being limited by a lack of intuitive visual analytics tools. Previous portfolio visualization systems have mainly focused on the relationship between the portfolio return and stock holdings, which is insufficient for making actionable insights or understanding market trends. In this paper, we present \textit{sPortfolio}, which, to the best of our knowledge, is the first visualization that attempts to explore the factor investment area. In particular, \textit{sPortfolio} provides a holistic overview of the factor data and aims to facilitate the analysis at three different levels: a Risk-Factor level, for a general market situation analysis; a Multiple-Portfolio level, for understanding the portfolio strategies; and a Single-Portfolio level, for investigating detailed operations. The system's effectiveness and usability are demonstrated through three case studies. The system has passed its pilot study and is soon to be deployed in industry. }


\CCScatlist{ 
 \CCScat{K.6.1}{Management of Computing and Information Systems}%
{Project and People Management}{Life Cycle};
 \CCScat{K.7.m}{The Computing Profession}{Miscellaneous}{Ethics}
}

\teaser{
  \centering
  \includegraphics[width=\linewidth]{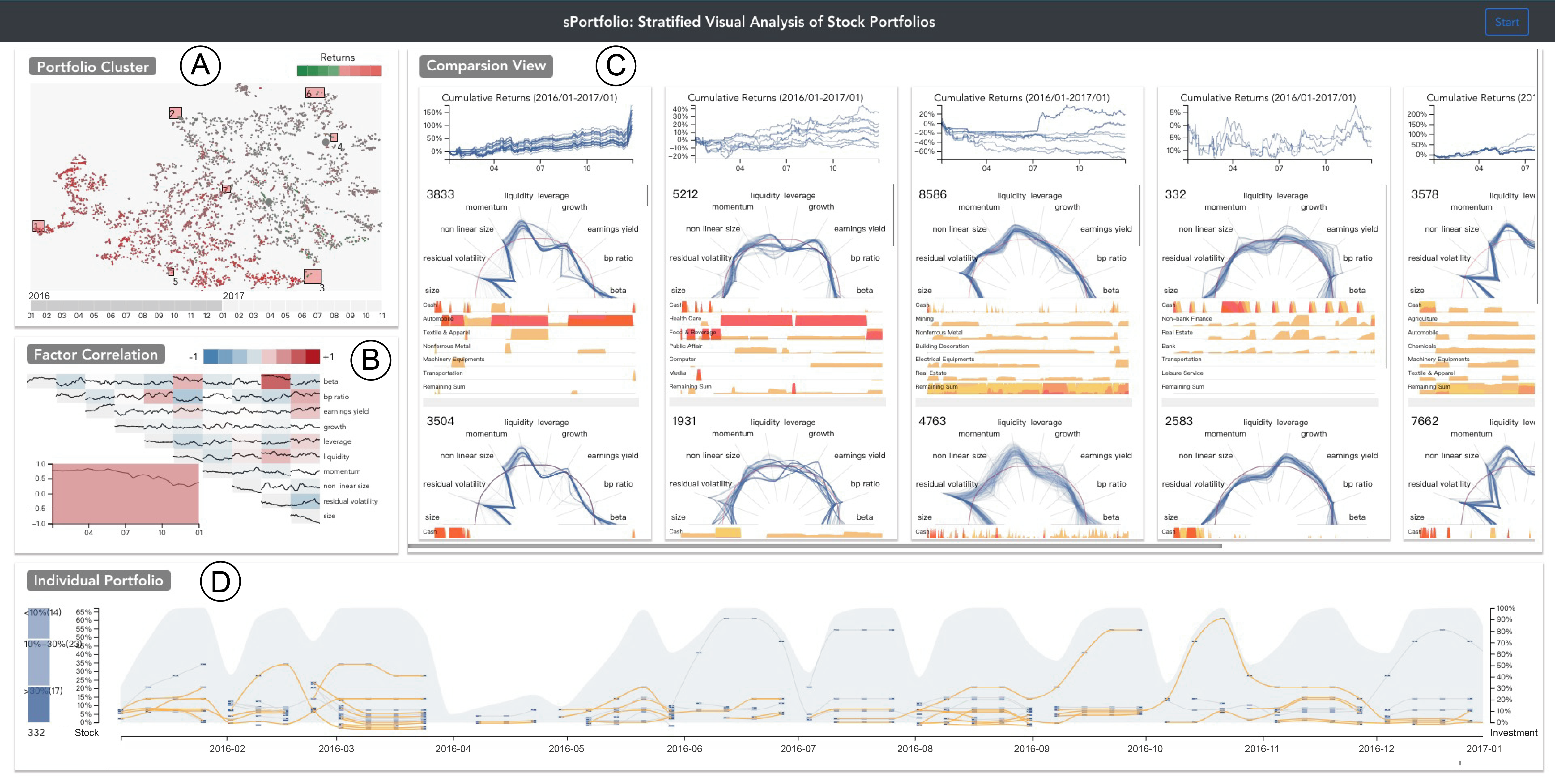}
  \caption{With \textit{sPortfolio}, we can observe the management and strategy of stock portfolios from different perspectives. A) The portfolio cluster view gives an overview of all stock portfolios within a given time-period. B) The factor correlation view reveals the return correlations of risk factors, which can be used to validate the effectiveness of the factors in a multi-factor model. C) The comparison view is designed to compare the risk preference and sector position of portfolios, which reveals their strategies. D) The individual portfolio view illustrates the stock holdings and trading style of a specific portfolio.}
  \label{fig:teaser}
}

\vgtcinsertpkg

\begin{document}



\maketitle

\section{Introduction}
With the development of computer technology, quantitative strategies play an increasingly important role in investment. The volume of quantitative-based investment is significant. At the end of 2017, 930 billion USD were being managed in funds, using related strategies. Moreover, since 2007, there has been an overall annual growth rate of 6.4\% \cite{hedgefund2018}. Risk factor analysis, i.e., where the risk factors concerning stocks or other instruments are calculated and analyzed, lies at the core of quantitative investment. The results of this analysis can then be used to guide investment decisions directly. However, the tools that are currently available for the research and analysis consist only of generic data analysis software, which does not cater to the needs described in this context. In this paper, we propose a visual analytics system that boosts the efficiency of factor research and portfolio analysis.

Over the years, researchers have proposed thousands of factor strategies over the years and have constructed backtesting portfolios that use past market data, in order to validate the effectiveness of the factor models and portfolio strategies. In doing so, vast amounts of valuable data from the portfolios are generated but underutilized, because of the lack of efficient tools to manage it. Almost every country's domestic stock market contains thousands of stocks, and each stock contains approximately thousands of dimensions of factor data per year. Moreover, the hyper-dimension nature of factor data creates a high barrier that blocks most investors from conducting research on factors and portfolios by traditional visual representations (line graphs or other basic charts). All of the above makes using traditional tools, to carry out portfolio analysis and factor research, extremely inefficient. Our proposed techniques perform three important tasks that help factor investors with their day-to-day analysis and boost efficiency. Firstly, in order to define the risk factors of financial instruments more accurately in this paper, we adopt one of the most innovative and popular factor models, the Barra risk model. This model codifies 40 different risk factors of a stock which can be quantitatively measured and tracked\cite{grinold2000active}. However, the Barra factor model has limitations. For example, it does not allow investors to examine the effectiveness of different factors at different times, which could help investors analyze the market situation over time. Secondly, investors need to study the past risk preferences of portfolios and to categorize the portfolios into different groups for further research. Lastly, after investors have decided which portfolios they are interested in, they may wish to look further into the specific industry holdings and trading strategies. 

To address the above challenges, we propose \textit{sPortfolio}, a comprehensive visual analytics system, which allows users to observe the market of portfolios in terms of their composition, factors, and historical strategies. We use the Barra risk model, the most well-known and predominant model used in factor management, to calculate the daily factors. \textit{sPortfolio} offers four well-coordinated views: the portfolio cluster, the factor correlation, the comparison, and the individual portfolio views, as shown in (Fig.\ref{fig:teaser}). Our system allows users to derive critical insights about the importance of, and the relationship between factors, and to discover the strategies used by other portfolios. This information can then be used to create new strategies and to make decisions about the construction of their own portfolios. To prove the efficiency of \textit{sPortfolio}, we conducted three case studies with domain experts from several financial institutes, which serves as a pilot study for commercializing our system in the near future.

In summary, the major contributions of this paper are as follows:
\begin{itemize}
    \item A visual analytics method of stratifying the stock portfolio data and visualizing the strata effectively.
    \item A visualization system, which allows users to explore and compare the detailed performance and management of stock portfolios.
    \item A set of comprehensive case studies with our system in collaboration with experts from various investment banks and quantitative institutions.
    
\end{itemize}

\section{Related Work}

\subsection{Factor Investing}

Quantitative investment began roughly in the 1960s when the Capital Asset Pricing Model (CAPM)\cite{sharpe1977capital} was invented. Since then various theories have been proposed to explain stock returns, using different macro and company-specific risk factors. Today, a great number of investors are able to research and create portfolio strategies based on risk factors due to technology advancement. According to the IPE report in 2017\cite{IPE}, the value of assets, under management by factor-related portfolio strategies, reached \$559 billion by the end of 2016. 

Factor investment adopts mathematical methods to construct portfolios. Typically, in a conventional case, investors conduct in-depth research into companies by making on-site visits and by interacting with the leaders. However, factor investors do this by acquiring comprehensive data about the companies that are publicly available, such as the book-to-price ratio of companies. They then apply a factor model to find effective risk factors within the market. The risk factor exposure of a stock is the risk level of the stock relative to the market. In factor investing, risk factor models study the relationships between stock return and factor exposures\cite{defusco2015quantitative}, by running multivariate regression; including price-to-earnings, price-to-book, and so forth. After the multivariate regression, the returns of different factors should be independent of each other. In this way, investors construct their diversified portfolios by setting their preferred risk factor exposures.

As mentioned earlier, of all the factor models proposed previously, Barra Risk Model is the most widely-adopted model in the quantitative investment industry. It was created by Bar Rosenberg, the founder of Barra Inc., as discussed by Grinold and Kahn\cite{grinold2000active} in 2000. It has been frequently updated for different stock markets by MSCI\cite{MSCI}, who acquired Barra Inc. The reason for its popularity in the quantitative trading community is that not only does it consider the traditional factors proposed decades ago, but it also takes several new factors into account \cite{bender2013foundations}. For example, in the Barra Risk Model for the United States stock market, there are 13 style factors and 13 industry factors in total\cite{barra1998united}, which is more than the one factor in CAPM or the three factors in Fama-French model\cite{Fama1992}. With its large number of factors, the Barra Risk Model generates better empirical results and more accurate predictions for stock returns. For this reason, the portfolio factor data used in this paper are based on the Barra Risk Model, which serves as an example of portfolio factor data without losing its generality.

\subsection{Portfolio Visualization and Performance Visualization}

Portfolio data visualization has been a popular topic in the past two decades, for which many visualization systems and methods have been proposed. Typically, earlier research work visualized the return information and the stock positions, which were vital for measuring the performance of a portfolio. Matthias et al.\cite{schaefer2011novel} proposed a line chart with the background segmented and colored to encode return values. Ziegler et al.\cite{ziegler2010visual} used a “Pixel Bar Chart” to display and compare the return and volatility of stocks, industry sectors, and countries in real time. They both worked on utilizing the background of the line chart for extra information. Besides a line chart, many previous works were based on heatmap \cite{ alsakran2009visual, ziegler2008visual, ziegler2007relevance}, which were capable of demonstrating returns in various time periods, thus significantly increased the amount of information shown in the same amount of space. Clustering has also been applied for an efficient observation of market returns. Lei et al. \cite{lei2010visual} clustered all stocks into groups by their returns and displayed them in co-centered, consecutive rings. Xiong et al. \cite{xiong2002er} examined the returns of different geographical regions are concerned in mutual funds' portfolio construction. In addition, systems like FinVis\cite{rudolph2009finvis} or PortfolioCompare\cite{savikhin2011experimental} aimed to help non-expert in portfolio selection, based on expected returns. In order to display stock holdings, Jungmeister et al. \cite{jungmeister1992adapting} and Csallner et al.\cite{csallner2003fundexplorer} both implemented treemaps to tackle this task with improved flexibility of interaction. Dwyer et al.\cite{dwyer2002visualising,dwyer2003scalable,dwyer2004visualising} proposed a ``2.5 Dimensional'' graph, which plotted the clustering results of funds based on the stock holdings. 

However, to the best of our knowledge, no system exists for visualizing the style factor information of stock portfolios. The domain experts who collaborated with us also claim that no similar in-house system has been developed. The earlier work lacks an indispensable aspect of any stratified in-depth analysis of portfolios. Most of the previous visualization systems focused on either only the overall situation in the market \cite{ziegler2010visual, lei2010visual, alsakran2009visual, dwyer2004visualising, keim2006spectral, yue2019bitextract} or only analyzing particular portfolios \cite{csallner2003fundexplorer, jungmeister1992adapting}. However, in practice, both the factor research and portfolios analysis are of great importance for decision-making in investment. In that regard, we did not find any systems proposed yet.

\subsection{Financial Multivariate Time Series Data Visualization}

Financial data visualization is mostly about time series data. Not only the stock markets but also the banking and insurance industry are all covered by time series data. Lei et al.\cite{lei2011visual} proposed the concept of ``visual signatures'' for financial time series data. Sorenson et al.\cite{sorenson2013financial} developed a time series visualization in Bloomberg Inc. by combining the discrete glyph-based events, which assists in price fluctuation analysis. Additionally, there are outstanding surveys towards financial data visualization, like FinanceVis.net\cite{FinanceVis} and towards visual analysis\cite{EuroVis16Survey}.

In our scenario, the stock portfolio data is multivariate time series data, which is composed of daily factor data, sector data, and stock data. A similar data format also exists in most quantitative investment scenarios. There were different visualization techniques proposed for multivariate time series data in the visualization community. Generally, previous researches mainly contributed to efficient layout techniques\cite{peng2008method}, to detect trends and patterns\cite{hochheiser2004dynamic}, or to aimed at a specific area, such as clinical data\cite{shahar2006distributed}, bibliographic database\cite{chen2006citespace}, and so forth. Comprehensive surveys were also proposed, such as generalizing time-oriented data design in terms of visual analytics\cite{aigner2007visualizing}, and a series of operations performed on a conceptual space-time cube\cite{bach2014review}.

In the quantitative investment analysis, the time variable is different under the factor perspective and the stock positions perspective. The former emphasizes the detection of trends and anomalies, whereas the latter focuses on the time points of trading. However, as in Frank's taxonomy\cite{frank1998different}, the time axis is made up of two kinds of temporal primitives: namely, time points and time intervals, both of which fall under the financial data analysis scenario consideration. Also, Aigner et al. \cite{aigner2007visualizing, aigner2011visualization} suggested that time be treated as one quantitative variable among many others. Thus, we need to fulfill the tasks of stratified displaying and comparing simultaneously, within a limited 2D space. As far as we know, there is no specific design yet, which solves these requirements in the financial scenario.

\section{Background}

This section describes the background to factor investing in the China A-Shares Market, as well as six concrete tasks that were generated from collaborated companies and domain experts. Meanwhile, the table below explains some investment terms that are frequently used.

\begin{table}[!htbp]
\begin{center} 
\centering
\begin{tabular}{|p{0.21\linewidth}|p{0.69\linewidth}|} \hline

\multicolumn{1}{|c|}{\textbf{Term}} & \qquad\qquad\qquad \textbf{Explanation} \\ \hline

\multicolumn{1}{|c|}{Risk Factor} & A set of common factors that impact returns \\ \hline

\multicolumn{1}{|c|}{Factor Return} & The return attributable to a particular common factor \\ \hline

\multicolumn{1}{|c|}{\tabincell{c}{Factor \\ Exposure}} & Used to measure how much a stock or a portfolio is exposed to a certain risk factor \\ \hline

\multicolumn{1}{|c|}{\tabincell{c}{Factor \\ Investing}} & An investment strategy in which securities are chosen based on characteristics and attributes that may explain differences in returns \\ \hline

\multicolumn{1}{|c|}{Portfolio} & A portfolio is a grouping of financial assets such as stocks \\ \hline

\multicolumn{1}{|c|}{Backtesting} & The process of applying a trading strategy or analytical method to historical data to see how accurately it predicts actual results \\ \hline

\tabincell{c}{Value-weighted \\ Average} & A measure of security prices adjusted according to the market value of each security included in the average \\ \hline

\multicolumn{1}{|c|}{Sector} & A large segment of the economy \\ \hline

\end{tabular}
\end{center}
\caption{Investment terms used in the analysis of factor investing}
\label{Invest_Terms}
\end{table}

\vspace{-0.4cm}
\subsection{Data Abstraction}

\subsubsection{Data Overview}
Our visualized data is comprised of the backtesting records for each portfolio, the factor exposures of each stock, the sector categories of each stock, and the factor returns of the corresponding factors.

Our backtesting data consist of 8451 records of various strategies, contributed by quantitative investors on the RiceQuant\cite{RQ} platform. Each backtesting record represents a portfolio with intrinsic portfolio strategies. On any given day, a portfolio contains a list of stocks and the number of shares of corresponding stock held. The portfolios vary in length, as they are performed over different time ranges, from 2016 to 2018. Overall, the portfolios cover 99\% of the whole China A-Shares Market stocks. Due to the confidentiality issues, the names of the portfolios are replaced by the numbers ranging from 0001 to 8451.

The factor exposures of stock are the values that measure how much the stock exposes to a corresponding risk factor relative to the market. According to RiceQuant, the factor exposures of each stock are calculated from the financial statements of the corporation using their corresponding definitions, and then normalized within the market, to ensure that the market index has 0 exposures to all factors. Suppose $X$ is a factor exposure value calculated, and $ \overline{X}$ is the value-weighted cross-sectional average of $X$ value of all stocks. Meanwhile, $\sigma(X)$ is the standard deviation of $X$. Then the final exposure is calculated as $X' = \frac{X - \overline{X}}{\sigma(X)}$. For example, if a stock has a significant positive exposure of \textbf{size} factor, it means that the particular firm has a larger market capitalization than the other companies in the market. As a result, the return on the firm's stock can be explained partially by the fluctuation of \textbf{size} factor return. In order to aggregate the exposures of stocks in a portfolio, the factor exposure of a portfolio is measured by the value-weighted average of its member stocks. In addition to this, the industrial sector category of each stock is also included in our data set. The sector category indicates to which sector a stock belongs. There are 28 industrial sectors on the local stock market in total. We use one-hot encoding to indicate whether a stock belongs to a specific sector or not. Similarly, the industry position of each portfolio is aggregated by the value-weighted average of the corresponding one-hot encodings from the stocks that it holds. In summary, for each portfolio, there are 10 daily factor exposures and 28 sector positions from 2016 to 2018, and there are 8451 portfolios in total.

The factor return data consist of the daily factor returns of 10 selected factors introduced by the Barra China Equity Model 5 (CNE-5) \cite{CNE5}. The factor returns are the returns that are attributable to a certain factor. The return on an asset can be decomposed and expressed as the asset's exposure to the risk factors times the factor returns, and a firm-specific return. According to RiceQuant, the factor returns are obtained from the following multivariate regression model built on the stock market. In the following equation, the $ r^t_j $ is the return of stock j at time t, and the $X^t_{js}$ is the exposure of stock j on factor s at time t. Meanwhile, $f^t_s$ is the factor return of factor s at time t, and $u^t_j$ is the residual return of the stock j. 
\begin{equation}
    r^t_j = \sum_{s=1}^S X^t_{js} * f^t_s + u^t_j 
\end{equation}

The nature of factor return data is multivariate time series. We obtained the daily factor returns from 2016 to 2018 from RiceQuant. 

\subsubsection{Stock Market and CNE-5 Model}

The visualized stocks in our system are listed on the China A-Shares Market. A-Shares are the shares of China-based companies that are publicly listed in the Shenzhen Stock Exchange and the Shanghai Stock Exchange. Currently, they are only quoted in Renminbi(RMB) and are only available for mainland citizens and selected foreign institutional investors.
CSI 300 \cite{CSI300} and CSI 500 \cite{CSI500} are two important indices of A-Shares. CSI 300 is a capitalization-weighted index of the largest and most liquid 300 stocks in A-shares. It is used to measure the overall performance of the China A-shares Market. CSI 500 is an index that comprises the remaining largest 500 stocks after the CSI 300 have been excluded. It is used to indicate the small-to-middle market capitalization A-Shares. At the end of January 2018, the total market capitalization of A-Shares was around 17 trillion RMB, and there were around 3400 stocks being traded in the market.  
The CNE-5 Model was proposed in 2012 by MSCI, in order to reveal the dynamics of the China A-Shares Market and to help institutional investors for their investment process. On the CNE-5 model, 10 style-factors are proposed. These style-factors are \textbf{beta}, \textbf{momentum}, \textbf{earning yield}, \textbf{residual volatility}, \textbf{growth}, \textbf{book-to-price}, \textbf{leverage}, \textbf{liquidity}, and \textbf{non-linear size}.

\subsection{Task Analysis} \label{task_analysis}

During the past decades, factor research and portfolio analysis have been extremely important topics in the financial area. 
We aim to develop our system in a user-centric manner, using agile software development\cite{schwaber2002agile}, in order to fulfill the domain users’ requirements. The entire process lasted eight months and involved close collaboration with six experts. $E_A$ (a co-author of this paper), $E_B$, and $E_C$ are all three product managers from RiceQuant. The RiceQuant company has been providing quantitative trading-related services in mainland China for over four years, and its clients cover most of the Chinese banks and financial institutions. The three product managers have been working on the frontiers of factor investing and have gathered many of industry requirements. They are all eager to evaluate portfolios, such as portfolio management, etc. Another internal expert, $E_D$ (not a co-author), is a finance researcher, who has devoted themselves to factor model evaluation and development. Model effectiveness and the amount of factor crowdedness are the essential angles of model evaluation. $E_E$ (also not a co-author) is a senior fund manager who has been trading in the Asian markets via factor investing for ten years. Almost all of our traders avow that the portfolio strategy analysis could benefit them.

Therefore, we summarized a list of analytical tasks, following a series of structured interviews with domain experts and experienced market participants. We followed a typical user-centered design framework by using discussions, brainstorming, designing, prototyping, presenting, implementing and deploying. After several iterations, we collected their feedback and condensed it into a set of six primary questions. These are further classified into three levels as follows:

Two \textbf{Risk-Factor-Level} questions provide a general market sense to the investors with regard to the risk factors.

\begin{compactitem}
\item [\textbf{T.1}] \emph{What is the effectiveness of risk factor model's output over a certain time period?} The data we are visualizing comes from a model that is derived statistically. The validity of a risk factor model's output changes over time. This means a risk factor's return might be independent of other factors' at the beginning but strongly correlated after the market conditions evolve. A high correlation between factors returns indicates a weak model output. Updating an ineffective model could prevent investors from taking unnecessary risks and help them optimize the portfolio.
\item [\textbf{T.2}] \emph{What is the 'crowdedness' of each factor at any given time?} When capital flows into a single factor, the expected return on the factor may drop significantly because of the price appreciation. Thus, it is crucial for the investors to speculate hypothesize any future returns on factors before they make an investment and avoid factors that are “crowded”, even though they might have earned a high return in the past.
\end{compactitem}

Two \textbf{Multiple-Portfolio-Level} questions that let investors compare, understand and replicate past stock portfolio strategies.

\begin{compactitem}
\item [\textbf{T.3}] \emph{How do various groups of portfolio strategies evaluate risks and industries differently?} Portfolio strategies are created by human beings, thus, represent their risk and industry preferences. Investors who trust in small companies may take huge risks in the size factor while other investors who favor fast-growing stocks may carry more momentum risk. To understand the rationale behind portfolio strategy groups, investors need to analyze the risks taken and the industry held by the portfolios.
\item [\textbf{T.4}] \emph{How did previous quantitative investors implement the portfolio strategies?} As well as understanding the portfolio strategies, the investors also demand that strategies are quickly replicated and deployed to the market. Stock holdings within a timespan give investors hints about implementing portfolio strategies.
\end{compactitem}

Two \textbf{Single-Portfolio-Level} questions that help investors dive into the detailed operations of the fund or portfolio.

\begin{compactitem}
\item [\textbf{T.5}] \emph{Which trading style did the portfolios adopt?} Fund and/or portfolio managers adopt different trading styles in their practices. Managers may trade in high- or low-frequencies. They may also form a concentrated portfolio or highly-diversified portfolio. Understanding a single portfolio’s trading strategy gives investors insights into dimensions outside the risk factors.
\item [\textbf{T.6}] \emph{How to speculate the future return of a portfolio?} In the past, a star portfolio or fund could earn more than 100\% annually. Will this trend continue? The portfolio’s current holdings and the risks it is taking both have a big influence on its future returns.
\end{compactitem}

\section{System Pipeline}

\begin{figure}[h]
\includegraphics[width=0.8\linewidth]{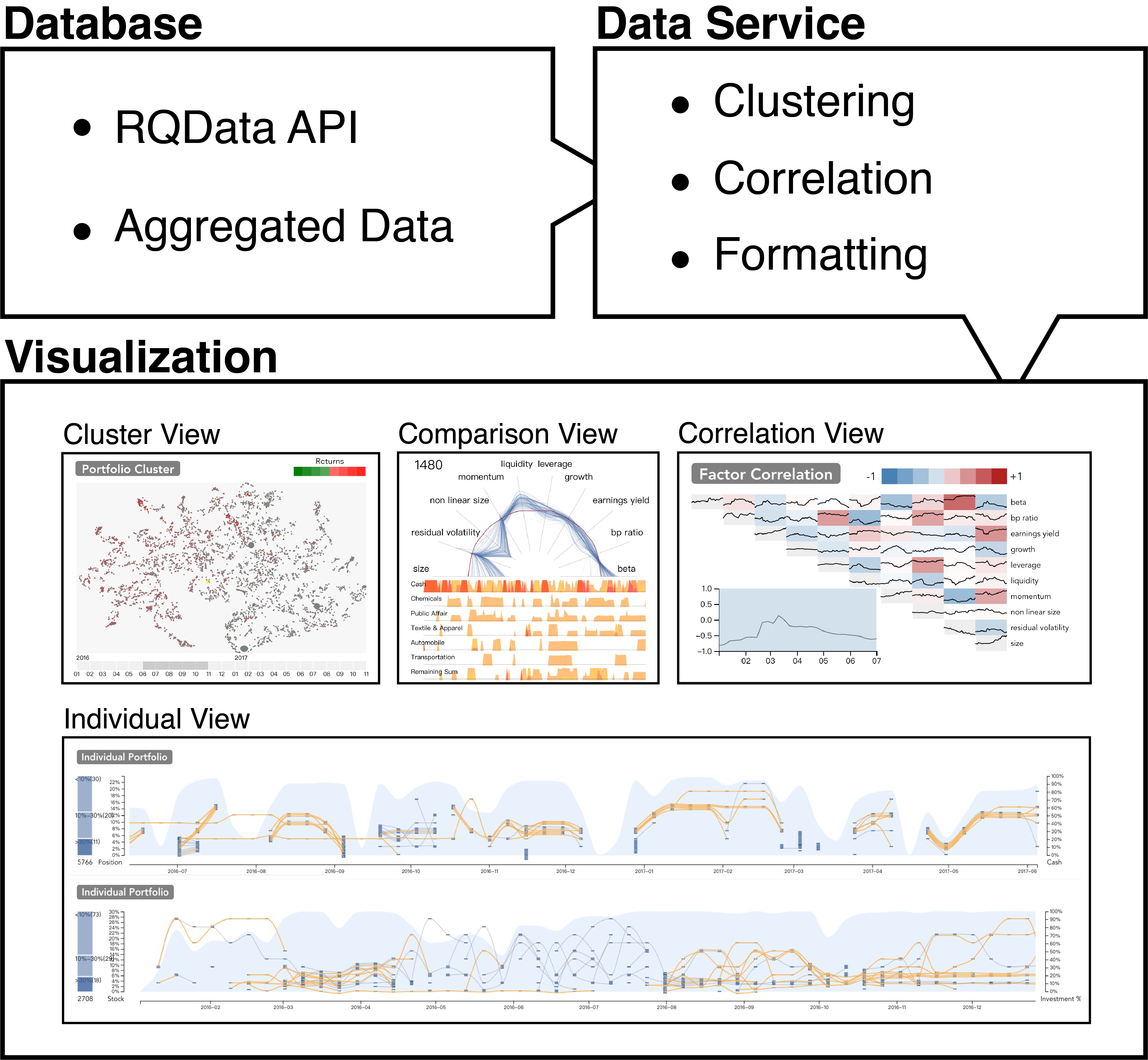}
\centering
\caption{System Overview. The three modules included in our system \textit{sPortfolio}: 1) Database; 2) Data Service; 3) Visualization, showing the detailed workflows inside each module.}
\label{pipeline}
\end{figure}

\textit{sPortfolio} is a web-based full stack application with three major modules, namely, a database module, a data service module, and a visualization module (as shown in Fig.\ref{pipeline}). The database module, which is based on MongoDB, holds two groups of information. One group is collected directly through the RQData API, which is provided by RiceQuant and includes details such as the daily stock factor exposures, the stock quantities and the respective prices of portfolios, as well as the total market value of a portfolio, every trading day. The other group is aggregated from the direct information gained from RQData. It contains calculated information, such as the portfolio sector positions, the factor exposures at the portfolio level, and the portfolio's returns. The data service module, which is based on Python Pandas and Tensorflow, further manipulates the data in the database to compute high-level information, such as the clustered results of the portfolios based on factor exposures and sector positions and the correlations between the factors over different time periods. Together, these two modules together serve as the back-end service and are hosted on a dedicated server, in order to boost up the overall responsive speed of the system.

The visualization module has four well-structured views that allow users to explore fully to obtain desirable insights. The portfolio cluster view shows all the portfolios after they have performed, clustered by their factor exposures or sector positions. The factor correlation view summarizes the returns and correlations of risk factors in the whole market. The comparison view allows users to select as many clusters of portfolios as possible for a quick comparison of investment styles between different clusters. The individual portfolio view examines portfolio management in more details.

In terms of the computational scalability, the data service module usually consumes the most time, throughout the whole pipeline of our system. However, it takes around 60 seconds for the data service to process thousands of portfolios and the visualization module to respond in a common browser (Chrome). All the statistics above are measured using a laptop with a four-core CPU and a built-in GPU. We expect to see a large time reduction once we have migrated to a high-performance computing server or cloud setting.

Here follows a general workflow of how \textit{sPortfolio} is used. The user begins by sliding the time window in the portfolio cluster view (Fig.\ref{fig:teaser}$A$), thus refreshing the clustering. Simultaneously, the correlations between the different factors and their returns are refreshed in the factor correlation view (Fig.\ref{fig:teaser}$B$) as well. If a user wants to see an enlarged view of the correlations and returns, they can hover the     mouse over the block, for a bigger plot in the factor correlation view. The user then brushes and selects some of the portfolios, displayed as dots in the portfolio cluster view, after which, the detailed information, including factor exposures and the industry holdings of the selected portfolios, are displayed in the comparison view (Fig.\ref{fig:teaser}$C$) as a portfolio overview. When a user clicks on the ID of the portfolio, in the comparison view, the stock holdings will be presented, over the entire life-cycle of the portfolio, in the individual portfolio view (Fig.\ref{fig:teaser}$D$).

\section{Data Model}\label{C5}

This section describes how we organize our data to achieve a satisfactory clustering result and how we derive the correlations between factors, from the market factor returns.

\subsection{Clustering}
The portfolio data is multivariate time series data of variant lengths because the portfolio is constructed within different time periods. For each trading day, every portfolio has a record of 39 dimensions including 10 factor exposures, 28 sector positions, and one cash portion. Thus, overall, we have 8451 portfolios whose life span vary from 20 to 400 trading days. We then face two problems, when we conduct an analysis over the portfolios. Firstly, the distance and similarity of the portfolios could not be measured directly, because they are not of the same length. Secondly, the number of features is relatively large, compared to the total number of samples, thus even if we use directly use padding and cutting techniques, the effectiveness and efficiency of the dimension reduction algorithms could not be ensured. To address these problems, we propose a pipeline, to reduce the dimensions, with two components, the first component is an LSTM autoencoder that maps the multivariate time series data of variant lengths into a vector in latent space. The dimensionality of the latent space should be neither too small, to encode the strategies behind each portfolio, nor too large, to be efficiently and effectively clustered. The second component is a dimension reduction algorithm that reduces the latent vectors to two-dimensional vectors, which are used as x and y coordinates in the portfolio cluster view.

\subsubsection{LSTM auto-encoder}
An auto-encoder is a type of neural network for learning the representations of data in a latent space \cite{ng2011sparse}, and which is widely applied to reduce the dimension of data.\cite{Hinton504}. Typically, it consists of two components, the encoder network $E$, and the decoder network $D$. Here we adopt the LSTM \cite{gers1999learning} structure for both. The latent representation of the i-th portfolio is obtained from the encoder network as $h^{(i)} = E(p^{(i)}, \theta_e )$, where the $\theta_e$ are the parameters of the enocder. Then the decoder network try to reconstruct the original sequence from the latent representation, which is $\hat{p}^{(i)} = D(h^{(i)}, \theta_d)$. In order to combine the encoder and the decoder network, the reconstructed sequence is expressed as $ \hat{p}^{(i)} = D(E(p^{(i)}, \theta_e) , \theta_d)$.
During the training process, we try to minimize the reconstruction loss, which is the Euclidean distance between the original sequence and the reconstructed sequence generated by the auto-encoder, by optimizing the parameters for the encoder and decoder network $\theta_e$ and $\theta_d$ simultaneously. Denote the optimized parameters are $\theta_e^*$ and $\theta_d^*$. Thus the optimized parameters could be expressed by
 
 \begin{equation}
     \theta_e^*, \theta_d^* = argmin_{\theta_e, \theta_d} \sum_{1}^{N} (p^{(i)} - D(E(p^{(i)}, \theta_e) , \theta_d))^2 
 \end{equation}
 The latent representation of each portfolio $p^{(i)}$, are calculated from the encoded network with optimized parameter $\theta_e^*$. The i-th latent representation is
 $ h^{(i)} = E(p^{(i)}, \theta_e^*) $. 
 Meanwhile, we could express the collection of latent representations as $H =  E(P, \theta_e^*) $.

We use Tensorflow 1.8 to implement LSTM auto-encoder. The number of hidden unit of LSTM is 50 and we use Adam-optimizer with a batch-size of 64. The model is trained on Nvidia 960m GPUs and the training time is around five hours. 

\subsubsection{Unsupervised dimensional reduction algorithms}
After using the encoder to encode the original portfolio data, we get the encoded representations of the portfolios $H$. Then we use dimension reduction techniques to further transform the representations $H$ into 2-dimensional vectors $C$. 
Here we choose t-SNE\cite{t-sne} algorithm because we are more concerned about the local structures of the portfolios. On the other hand, the t-SNE algorithm is able to maintain the stability of converging to a global optimum\cite{nonato2018multidimensional}. 
\begin{equation}
    C =  tSNE(H) 
\end{equation}
Where $C$ is the collection of $c^{(i)}$. Following that, we obtained the coordinate $c^{(i)}$ of each portfolio $p^{(i)}$ in order to display them in the portfolio cluster view. In the system, we use the t-SNE algorithm from the Python scikit-learn package.

\subsection{Factor Correlation}
The factor return is the return of a benchmark portfolio whose exposure to the corresponding factor is 1 and which has no exposure to other factors. We obtained the daily factor returns of the 10 style factors from 2016 to 2018. Then, we calculated the accumulated factor returns and the correlations among the factor returns.

\subsubsection{Accumulated Factor Return}
The accumulated return is an aggregated amount that is gained or lost over a certain time period. The accumulated returns can show factor trends more clearly than the original return. The expression of the accumulated factor return is expressed as $ R_j^i = \prod_1^i (r_j^k + 1) -1 $, where $r_j^k$ is the factor return of the j-th factor on the k-th day.

\subsubsection{Correlations of Factor Return}
We proposed two methods to measure the mutual correlations of all the factor returns. The first one is to measure the correlation of factors at a specific time. Denote $r_j^{n:m} = <r_j^{n}, r_j^{n+1}, ..., r_j^{m}>$. We calculate the correlation of factor $j$ and $k$ at day i by using the factor returns from 20 days before and after day i respectively. We choose 20 days because there are around 20 trading days in each month. Denote $corr$ as the function calculating the Pearson correlation between two vectors.
\begin{equation}
    \rho_{j,k}^i = corr( r_j^{i-20:i+20},  r_k^{i-20:i+20})
\end{equation}

The second method is a correlation of the selected time period. If we suppose s is the first day selected and t is the last day selected. The expression of that correlation is

\begin{equation}
    \rho_{j,k} = corr( r_j^{s:t},  r_k^{s:t})
\end{equation}

\section{Visual Design}

\textit{sPortfolio} consists of four components: the portfolio cluster view (Fig.\ref{fig:teaser}$A$), the factor correlation view (Fig.\ref{fig:teaser}$B$), the comparison view (Fig.\ref{fig:teaser}$C$) and the individual portfolio view (Fig.\ref{fig:teaser}$D$). We followed well-acknowledged design rationales to guide the design of these views, that we could ensure an effective delivery of visual information. 

Firstly, the design of the overall framework follows the Shneiderman’s mantra \textit{``Overview first, zoom and filter, then details on demand." }\cite{shneiderman2003eyes} Given the massive amount of multi-dimensional temporal data in the portfolio management, it is difficult to display the information on one screen, all at the same time. Therefore, the system presents an overview of portfolios first and then provides various interactions that allow users to filter the data and zoom-in for detailed information.

Furthermore, since the system provides various interactions, to show various data granularities of data, on user demand, it is crucial to arrange the information properly, so that users could perform analysis efficiently. According to the theory proposed in \cite{barsky2008cerebral}, it is more efficient for users to compare views side-by-side, rather than commit visible items to memory. For this reason, we attempt to use the screen space fully, in our system design, so that we could display different data granularities side-by-side. This avoids that users are forced to rely on a mental map to perform comparisons. 

Last, but not least, we design the system views and interactions to guide users, so they perform analysis intuitively and logically. The interactions of a single view and cross-view are designed to allow users to switch perspectives between different granularities easily so that they could examine data efficiently. 

A detailed description of the system is presented below. 

\subsection{Portfolio Cluster View}\label{cluster}
We provide a portfolio cluster view for users to observe similar portfolios efficiently. The market changes every day, which means different time periods may have different patterns and insights, such as different portfolio clustering results (\textbf{T3}). Users can gain a better understanding of the market via the portfolio cluster view. Additionally, the portfolio cluster view acts as the entrance of \textit{sPortfolio}; thousands of portfolios would overwhelm users. Different clusters with returns encoding could help the users find ``valuable'' clusters and outliers.

There are two parts in the portfolio cluster view, the clustering space and the timeline (Fig.\ref{fig:teaser}$A$). Based on the data model in Section \ref{C5}, we project all the portfolios to the clustering space and obtain coordinates $c^{(i)}$ for each portfolio $i$. Every node represents a portfolio with the color encoding its return. Under the China A-Shares market's current conventions, red color denotes a higher return and green color means a lower return. The two bigger nodes are the two most important benchmark portfolios in China A-Shares market; CSI 300 and CSI 500. The distance between the two nodes represents their similarity. Similar portfolios would gather closely and be combined into a small cluster.

The timeline embeds a brushing function which can triggers interaction in the portfolio cluster view (Fig.\ref{fig:teaser}$A$) and the factor correlation view (Fig.\ref{fig:teaser}$B$). After the time period is indicated after brushing, the data service module reperforms the clustering of portfolios and calculates the factor return correlations during the selected time period.

\subsection{Factor Correlation View}\label{correlation}

The factor correlation view (Fig.\ref{fig:teaser}$B$) shows the market performance of the risk factors, in terms of the cumulative market return of each factor and the correlations between these factor returns. This information helps the user to see the effectiveness of a factor model in a specified time period (\textbf{T1}), as well as any investment trends among fund managers for potential factor crowding (\textbf{T2}), which may affect the return of a portfolio in the near future (\textbf{T6}).

There are three parts to the information in the factor correlation view, as shown in Fig.\ref{fig:teaser}$B$. They are the upper right part, the diagonal part, and the lower left part. The design of the upper right part is built on a heat map. The x- and y-axes of the heat map encode the factor types, from right to left and from top to bottom, respectively in the same order. The color of each block, in the heat map, encodes the correlation between the corresponding two factors, which is the $\rho_{j, k}$ in Section \ref{C5}, in the selected time period from the timeline of cluster view (Fig.\ref{fig:teaser}$A$). The blue and red color in a block stands for a negative and positive correlation between the corresponding factor returns respectively. Simultaneously, we use the color's saturation to encode the magnitudes of the absolute values of the correlations. Inside each block, there is a line chart indicating the trends of the correlations in greater details. The coordinate of the i-th position in the line chart is $\rho^i_{j, k}$ as explained in Section \ref{C5}. The vertical scale of all the lines is unified across the view. 

For the diagonal part (in the dotted rectangles), the background of each block is always gray, and the line charts in the blocks show the corresponding cumulative market factor returns $R^i_j$ from Section \ref{C5}. 

The lower left part is used for displaying the details of the line charts. It is linked to the other two parts through on-hover mouse events. When a mouse hovers over a particular block, the corresponding line chart within the block, together with the background color will be displayed in the lower left-hand part. Additionally, we use the y-axis to encode the exact amount and the x-axis to encode the corresponding timeline.

\textbf{Justification: }We separate the view into three components because we want to observe the factor returns and their correlations at different scales. An alternative design would be to apply a fish-eye view to the line chart, instead of using colors to encode the larger time scale information. A fish-eye view\cite{sarkar1992graphical} is often used to display global context and local details simultaneously; however, there are some drawbacks to this alternative design. Firstly, the fish-eye distortion over the y-axis, which encodes the time information, may confuse the user. Secondly, we are more concerned with the larger time-scale performance of the factor return, based on the domain's requirements. Lastly, the fish-eye view does not fit into a global context, intuitively. This is why we choose to use the color channel to encode the information in a larger time-scale and the line chart to show the local trends.

\subsection{Comparison View}\label{comparison}
The comparison view provides overviews of various portfolios, which reveals the general patterns in the portfolios, at a glance (\textbf{T3}) and enabling users to compare risk and industry preference quickly between portfolios (\textbf{T4}). Together with the portfolio cluster view and the factor correlation view, the comparison view also stimulates speculations about the portfolio's return in the near future (\textbf{T6}).

\begin{figure}[h]
\includegraphics[width=1\linewidth]{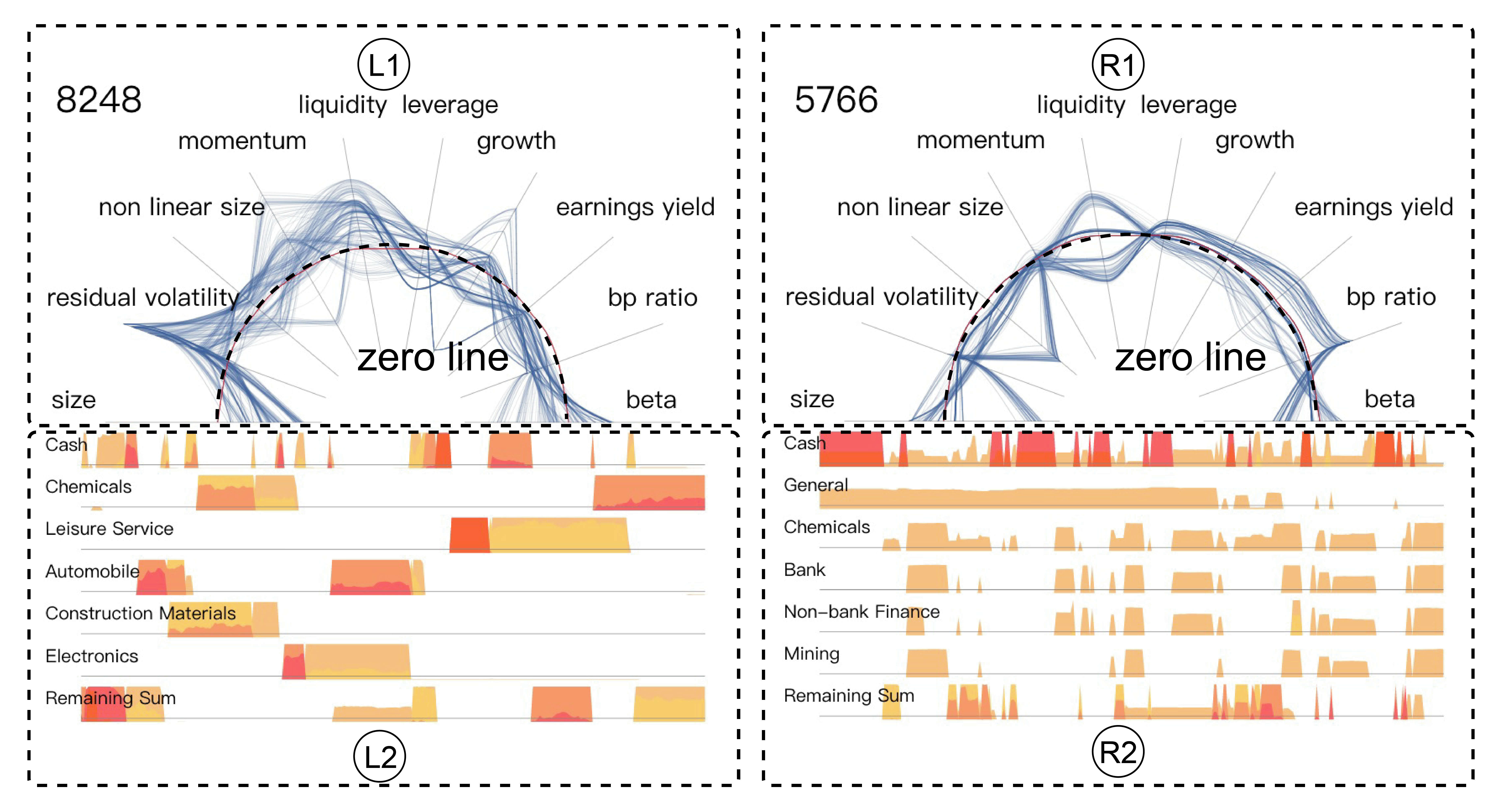}
\centering
\caption{Overview of two portfolios. The upper parts are factor signatures (L1 and R1) and the lower parts are sector graphs (L2 and R2). The blue lines, in L1, are sparsely distributed and do not converge on any axis. This means that the portfolio does not have a consistent risk preference. 
On the right side, the lines in R1 are densely distributed and often converge to one to two points where they pass through the axes. This means the factor exposures of the portfolio are less fluctuated, indicating a clear factor strategy. The horizons of the sector graph have longer segments in L2. Conversely, in R2 there are many short segments with many breaks in between. These observations show that the portfolio 8248 is managed by sector strategy while 5766 by factor strategy.}
\label{glyph}
\end{figure}

\textbf{Description: }As shown in Fig.\ref{fig:teaser}$C$, the comparison view is divided into regions that are horizontally juxtaposed to each other and that extend from left to right. Each region is dedicated to one cluster of portfolios, which is formerly selected in the portfolio cluster view. Within a region, the cumulative returns of all the portfolios are displayed at the top, followed by \textit{overviews} of each portfolio in the cluster one-by-one. Note that, as the result of clustering, the portfolios inside a region should exhibit similar risk and industry preference.

Fig.\ref{glyph} shows an overview of a portfolio, which consists of two major parts: the \textit{factor signature} as the upper part (Fig.\ref{glyph}$L1$) and \textit{sector graph} as the lower part (Fig.\ref{glyph}$L2$). There are ten co-centrical circular axes, in the factor signature, representing the factor exposure values of the ten risk factors in the CNE-5 model. A line appears in the factor signature and is drawn by connecting the points projected to the ten axes by the ten factor exposure values of the portfolio every trading day. The larger the value, the more distant it is from the common center. The color in the signature is darker where the lines overlap. The factor signature is dedicated to identifying the range of fluctuations of factor exposures for each factor as shown in Fig.\ref{glyph}. 

The sector graph is a horizon graph\cite{horizongraph} that encodes seven portions of a portfolio including cash, the top five sectors of the holding, and the sum of the rest sectors in a top-down manner. Time elapses as the graph continues from left to right. The vertical level and the color of the horizon line indicate the amount, which is typical of a horizon graph.

\begin{figure*}
    \setlength{\abovecaptionskip}{0pt} 
    \setlength{\belowcaptionskip}{0pt} 
    \centering
    \includegraphics[width=1\linewidth]{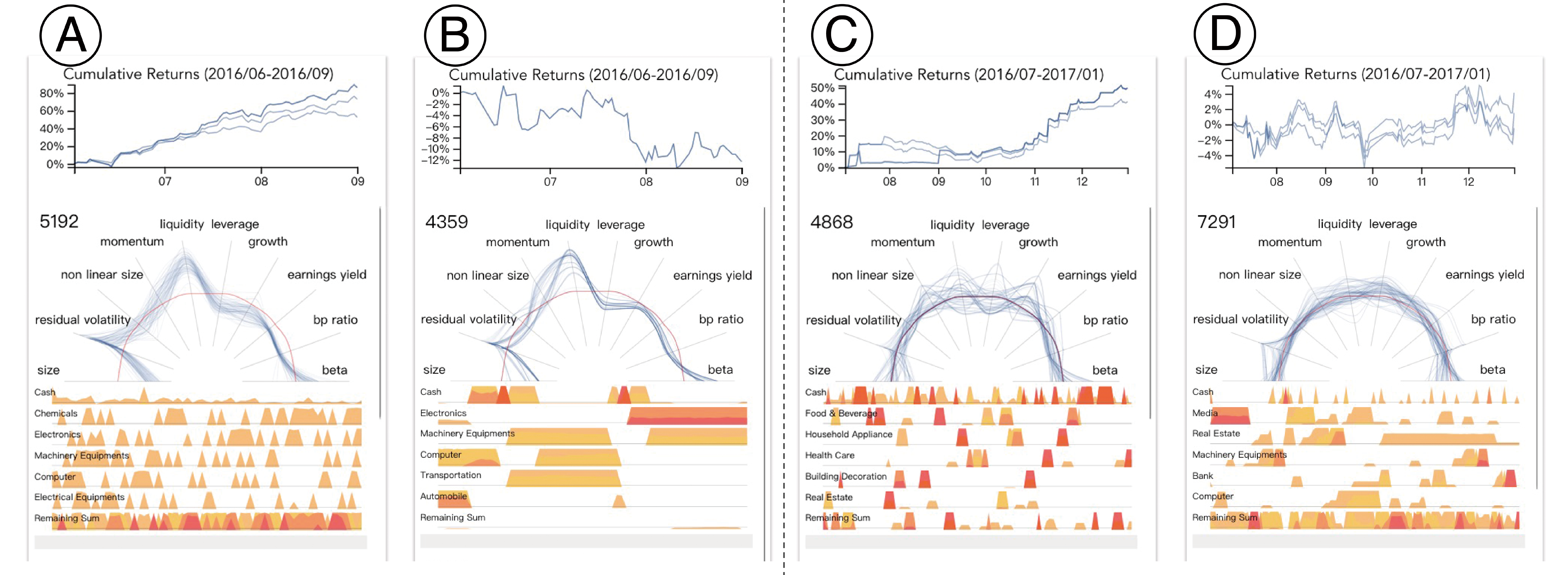}
    \caption{Various portfolios displayed in the comparison view. The line chart on the top of each cluster shows the cumulative returns in the selected cluster. The portfolio overviews (5192, 4359, 4868, 7291) are exhibited in two different periods. This shows that, 5192 and 4359 had very similar risk factor exposures as a group during 2016/06-2016.09. The same was true for 4868 and 7291 during 2016/07-2017/01. Finally, their returns differed drastically from each other within the group. Meanwhile, within the groups, the portfolios had very different industry holdings.}
    \vspace{-0.2cm}
    \label{C1_1}
\end{figure*}

\textbf{Justification: }The overview is split into two parts because there are two separate sorts of information concerning a portfolio: the factor exposures, and the sector positions. Parallel coordinates are used in the design of the factor signature, which deliberately omits the time dimension. This is because, after the time period that is already specified in the cluster view (Fig.\ref{fig:teaser}$A$), the distribution of the data is the greatest priority for illustrating the style of a portfolio (\textbf{T3}). Within this context, the exact timing of a particular data (for example an outlier) is not a concern and is sometimes prone to noise, according to our domain expert. Another possible design would be to stretch the axes to be placed horizontally, However, the users tend to compare the data vertically rather than focus on the entire picture.
Our adopted design used circular axes which makes it easier for the users to perceive the graph as a whole signature, aiding speedy comparisons. 

We have tried a stacked area chart for the sector graph. However, because it is not unusual for stocks to be purchased in and sold out repetitively, a dedicated area in a stacked area chart for a sector would constantly appear and disappear, making it challenging to follow the changes and trends, simply. Additionally, the potential color choices would quickly run out of more sectors were concerned. 

The overall design of the comparison view is highly space efficient, to facilitate various portfolios' being displayed in a single view and to allow the quick identification of discrepancies between portfolios in multiple aspects, including the return, factor, and sector.

\subsection{Individual Portfolio View}\label{individual}

The individual portfolio view displays the management details of a portfolio, at the stock holding level, which helps the users further examine and confirm the portfolio's strategy (\textbf{T4}) and the style in which the portfolio is managed (\textbf{T5}).

\textbf{Description: }In the individual portfolio view ( Fig.\ref{fig:teaser}$D$), there is one horizontal axis that encodes the timeline and two vertical axes, representing the stock percentages and the total percentage of investment, respectively. Note that there are two sets of information in the graph: the background and the foreground. In the background, the height of the theme river represents the portion of the total investment. It also corresponds to the vertical axis on the right-hand side.
 
In the foreground, there are multiple horizontal sticks stacked from bottom up, at each timestamp, each of them represents a particular stock. Sticks that are overall close to each other, overall, form a minor group and the stocks that fall into the same group take up a similar number of percentages within a range of minor-extent. There are, at maximum, five groups of stocks, at each timestamp, with ranges, which are evenly distributed from 0 to the largest percentage a stock has ever taken up. The exact amount can be inferred from the vertical axis on the left-hand side. The lines connecting the sticks denote those stocks that are continuously in holding and will stop when the stocks are all sold out. Once an on-hover mouse event is triggered, the color of these lines will be changed to red, which is eye-catching. Simultaneously,
the order book id of the corresponding stock will be displayed near the mouse position. As the stocks are being bought in and sold out throughout the whole life span of the portfolios, the lines in the individual portfolio view fluctuate accordingly, indicating the trading frequency of the portfolio.

A stacked bar chart is displayed to the far left as an interactive legend. Once selected, each bar will trigger an event that highlights the lines of those stocks that satisfy specific criteria; namely, the period of the stock holding \textgreater30\%, belongs to 10-30\%, or \textless10\% of the overall lifetime of the portfolio, respectively. 

\textbf{Justification:}
We considered multiple alternative designs including line charts and stacked area charts. However, as the number of stocks involved increases, the cognitive load of the graph rises drastically. The line chart works well when only a few stocks are drawn, typically $\leq$ 5 stocks. In the case of a common portfolio, and based on our data, when 10-60 stocks are concerned, there are too many line-crossings and visual clutters for a human to digest. In a stacked area chart, the area for a stock is hard to track, when the stock is purchased and sold out repetitively and may lead to a shortage of color choices where more stocks are involved.
Our design for the individual portfolio view is compatible with more stocks being shown and still manages to convey the information clearly.

\subsection{Cross-view Interaction}

In addition to the interactions within the individual portfolio view which are discussed above, the system also supports interactions across different views, to help users perform efficient analyses. 

\textbf{Temporal-related Interactions. }These interactions enable users to focus on analyzing portfolios within a given period. By default, our system displays all portfolios in an arrange from 2016 to 2018. The user can zoom in to any period by brushing the timeline in the portfolio cluster View. After brushing, the portfolio cluster view and the factor correlation view will be updated accordingly. Users can then proceed to the following interaction for detailed analysis (\textbf{T1, T2}).

\textbf{Progressive zoom-in for drill-down analysis. }The interactions guide users to perform a thorough analysis of portfolios of interest. In the portfolio cluster view, users can select the clusters of interest, which puts overviews of the various portfolios into the comparison view to show the general patterns of portfolios (\textbf{T3}) for quick comparison (\textbf{T4}). Users can then click on a specific portfolio in the comparison view, to see the portfolio's strategies in the individual portfolio view, which shows the stock holdings of the portfolio. The level of details helps users examine the portfolio's strategy (\textbf{T4}) as well as its trading style (\textbf{T5}). It should be noted that the individual portfolio view shows the stock holding conditions over the entire span from 2016 to 2018, and the corresponding node of the portfolio in the portfolio cluster view (Fig.\ref{fig:teaser}$A$) is annotated by a yellow circle. The interactions crossing from the portfolio cluster view to the individual portfolio view guide users to zoom-in progressively from the entire pool of portfolios to the clusters of interest, and eventually all the way down to the detailed management of a specific portfolio, in order to perform a drill-down analysis. 

\begin{figure*}
    \setlength{\abovecaptionskip}{0pt} 
    \setlength{\belowcaptionskip}{0pt} 
    \centering
    \includegraphics[width=1\linewidth]{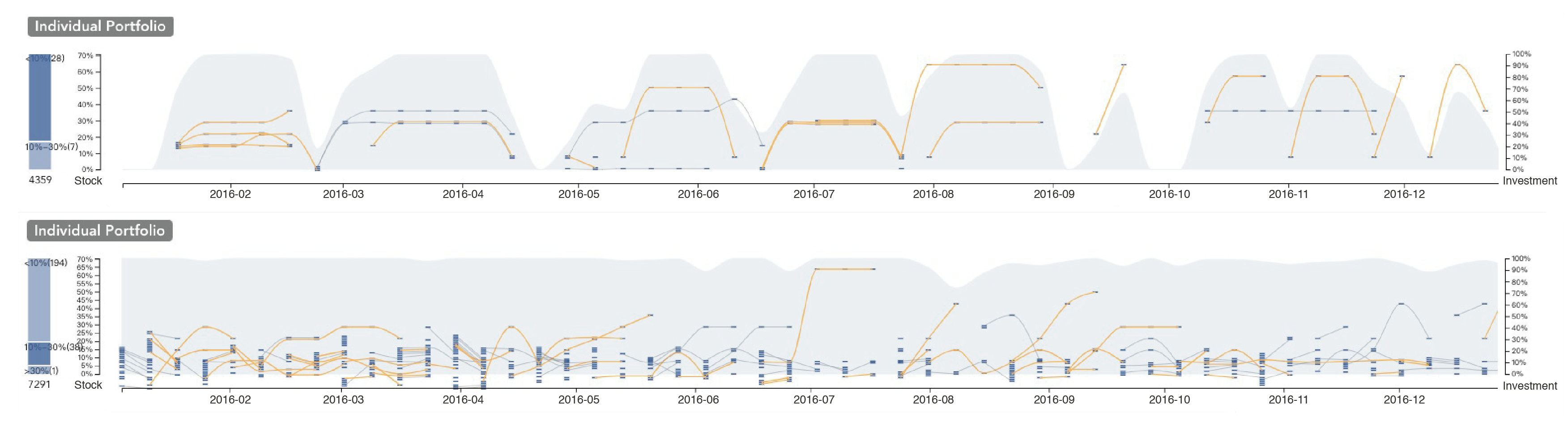}
    \vspace{-1.5em}
    \caption{Stock position details for the portfolios in Case 1, including 4359 (top) and 7291 (bottom). The vertical levels of points on the lines of 5192 do not vary so frequently comparing to the case of 7291, which shows that 7291 possessed a higher trading frequency. More sticks appear in 7291, indicating that more stocks are involved in the portfolio management.}
    \vspace{-0.2cm}
    \label{C1_2}
\end{figure*}

\section{Evaluation} \label{case}

The section below presents the three case studies in detail. The \textit{sPortfolio} system is designed to provide a toolset for investors to use to efficiently compare and analyze thousands of portfolios and thereafter to better understand different the various market conditions. To evaluate the system objectively, we conducted three case studies with a fund manager ($E_E$), a quantitative researcher ($E_D$) and product managers from RiceQuant ($E_A$, $E_B$, $E_C$).

\subsection{Multiple Portfolio Comparison}

When investors look back and reflect on their own and other portfolios, they look for the reasons behind the shining results of a star portfolio or the poor performance of other portfolios. Our fund management expert, $E_E$ planned to compare portfolios that outperformed and underperformed the market during a specific period and hoped to gain some insights (\textbf{T3}). In this case, the expert first wanted to classify thousands of portfolios into groups and then to pick the groups he wanted to probe into more deeply (\textbf{T3}). Following that he compared the groups of portfolios in terms of their strategies (\textbf{T4}) and trading styles (\textbf{T5}). 

$E_E$ began by selecting four clusters from two periods of time in the portfolio cluster view, and compared them, two-by-two, in the comparison view. As shown in Fig.\ref{C1_1}, the two portfolios on the left and the other two portfolios on the right were selected from the same time period. $E_E$ found that within the two groups, the portfolios had similar strategies, in terms of factor investing, as their factor signatures were similar. The factor signature is a visual design introduced in Sec.\ref{comparison} and Fig.\ref{glyph}. From the factor signatures, $E_E$ noticed that most lines bound inward significantly on the \textbf{size} factor and outward on the \textbf{liquidity} factor. This meant that many stocks that had small market capitalization and high liquidity. The lines in the factor signatures of the other two portfolios were close to the zero lines on almost every factor, which represents these portfolios were very close to the market portfolio. He then clicked on the portfolio IDs from within the comparison view. As shown in the individual portfolio view in Fig.\ref{C1_2}, the lines in the first two graphs did not move up and down a great deal, and the stocks had a major proportion in the portfolio. The lines in the other two portfolios did move up and down all the time and the stocks usually took up no more than 5\% of the portfolio weight. These observations convinced him that the first two portfolios adopted a low-frequency trading approach while the other two portfolios adopted a high-frequency trading style.

Interestingly, the expert found that in the aggregated cumulative return graph, there was a portfolio that generated a return of more than 50\% and another portfolio, within the left group or the right group, which had only a small return or even suffered a loss. The portfolios within the same group had the same or similar factor signatures shown by the shapes of the lines, but they had significantly different returns during the time periods. $E_E$ was curious about the reason behind the differences. He looked more closely into the comparison view as shown in Fig.\ref{C1_1}, and discovered that the portfolios within the same group were using different industry strategies, even though they had similar risk exposures. Neither the name nor the ranking of the industries of the portfolios was similar in the comparison view. $E_E$ was finally persuaded by the system that industry strategy was the reason behind the contrasting returns.

\subsection{Single Portfolio Performance and Factor Crowdedness}

Similar to stocks, factors that enjoyed superior returns in the past are likely to attract more capital, which pushes the returns of the factor even higher. However, the expected return of any financial instrument is limited, including its factors. We assume factors with superior returns over long periods to have a high level of crowdedness and therefore believe they would suffer losses in the near future (\textbf{T2}).
This is because crowded factors are more likely to have drawdowns than uncrowded factors \cite{factorCrowding}. If a portfolio is chasing a factor with a rising return, the return will be unsustainable since the crowdedness of the factor becomes increasingly high (\textbf{T6}). Therefore, factor crowdedness is a good indicator of how to speculate on a portfolio's future return.

\begin{figure}[h]
\includegraphics[width=1\linewidth]{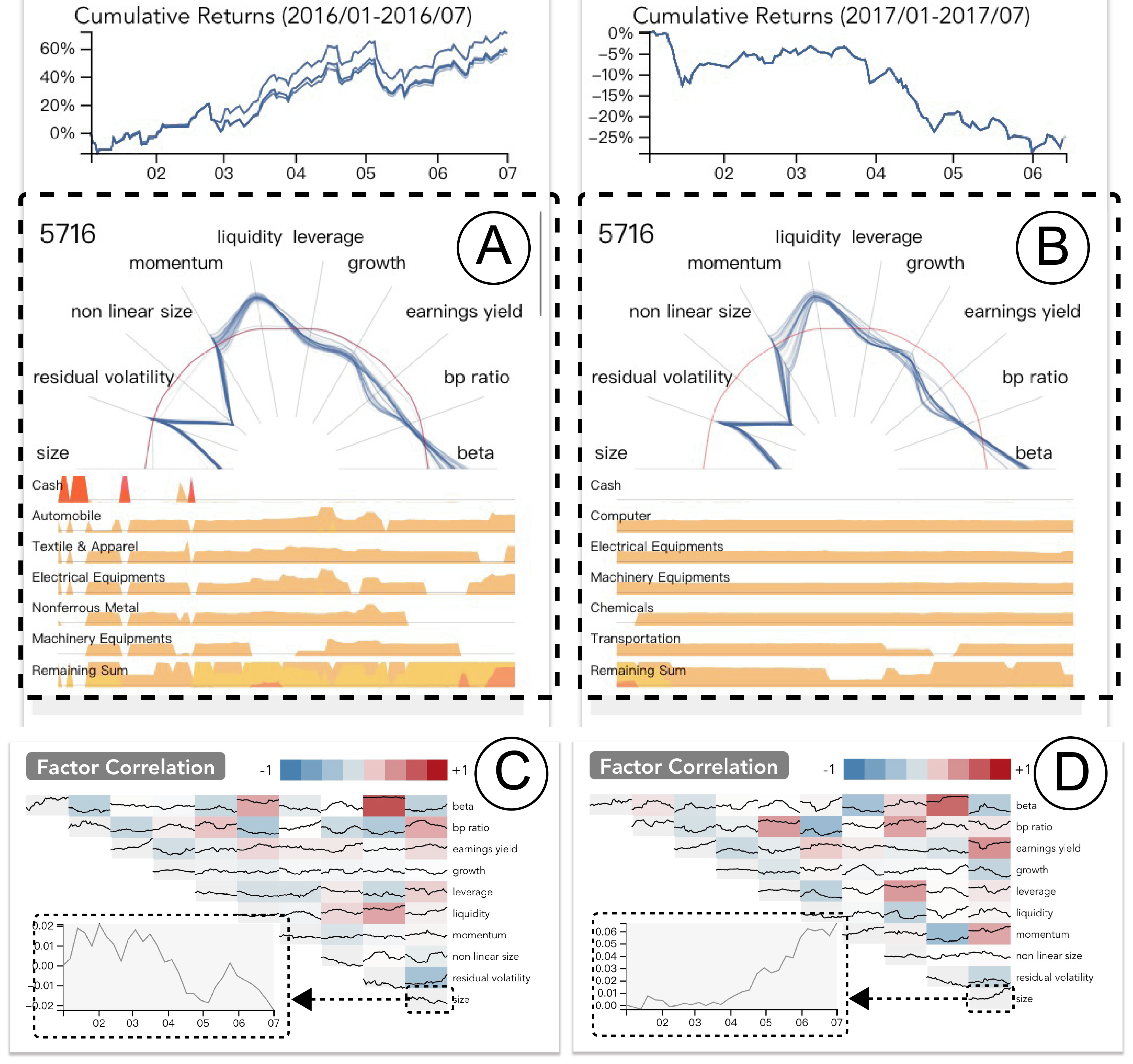}
\vspace{-1.5em}
\centering
\caption{Performance and factor return of the portfolio 5716 during different time periods. A) The portfolio took high \textbf{size} and \textbf{non-linear size} risks during 2016/01-2016/07. The portfolio enjoyed high return. B) The same portfolio (5716) maintained a similar strategy (factor and sector position) during 2017/01-2017/07, whereas the returns performed adversely. C) The \textbf{size} factor's return dropped during 2016/01-2016/07, which meant the portfolios that took negative \textbf{size} factor exposure would have benefited. D) The \textbf{size} factor's return was raised during 2017/01-2017/07, which meant the portfolio that took positive \textbf{size} factor exposure would have suffered from loss. The same strategy may perform totally differently under different market situations and factor crowdedness.}
\label{C2}
\end{figure}

$E_D$ explored the portfolio data with \textit{sPortfolio}, hoping to find some evidence of high factor crowdedness via different views (\textbf{T2}). First, he slid the time window, in the portfolio cluster view, to the first half of 2016 to cluster the portfolios into groups according to their factor exposure in that period. Several portfolios stood out among all the clusters, with a bright red color that indicated superior performance relative to the benchmark return. $E_D$ then brushed and selected the portfolios to check which factors were behind the superior returns (\textbf{T4}). The factor signatures of the selected portfolios in Fig.\ref{C2}$A$ showed identical risk preference and strategy. Most of them preferred \textbf{liquidity}, \textbf{non-linear size}, \textbf{book-to-price ratio} and \textbf{size} risk factors and demonstrated the four risks significantly. The lines in their factor signatures deviated from the zero line significantly on these four axes. 

To validate the effect of the four risk factors on the returns of those portfolios, $E_D$ went to the factor correlation view in Fig.\ref{C2}$C$ to check the cumulative returns of the four factors in the first half of 2016 (\textbf{T4}). \textbf{Size} and \textbf{non-linear size} factor had continuous negative returns whereas the other two factors did not fluctuate greatly. In this case, negative \textbf{size} and \textbf{non-linear size} exposures generated a positive return because the risk returns were negative. $E_D$ then suspected that these two risk factors had high factor crowdedness during the selected period (\textbf{T2}). To verify his hypothesis, he slid the time window in the portfolio cluster view to one year later, the first half of 2017, and then clicked on the ID of the portfolio, he had selected before in the comparison view to highlight it. To check the returns of the same portfolio in the first half of 2017, $E_D$ brushed and selected the highlighted portfolio in the portfolio cluster view. In the factor correlation view in Fig.\ref{C2}$D$, he observed that the trend of the \textbf{size} factor return was reversed in the later time period but the \textbf{non-linear size} factor did not change. The same portfolio, which had demonstrated the four risk factors in the first half of 2016, still had the same risk preference (\textbf{T4}) in early 2017 as it shows in Fig.\ref{C2}$B$, where their factor signatures look quite similar. However, in the second time period, the portfolios had a negative return because the portfolio managers didn't anticipate high factor crowdedness and speculated the factor returns to be quite promising. $E_D$ then concluded that the \textbf{size} factor was indeed having very high factor crowdedness in mid-2016. 

\begin{figure}[h]
\includegraphics[width=1\linewidth]{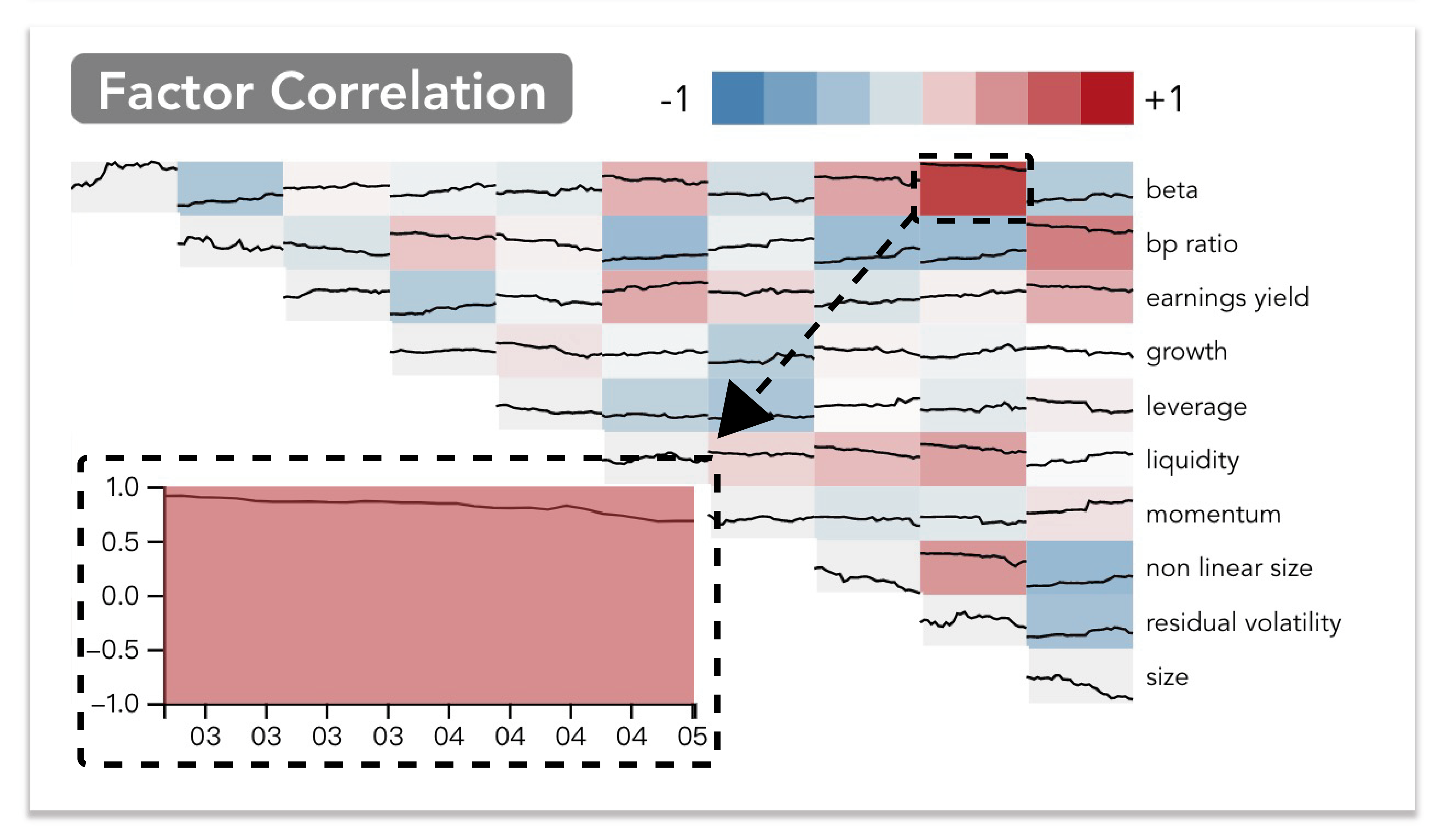}
\vspace{-1.5em}
\centering
\caption{Factor correlation view. The return correlations of most factors were highly positively or negatively correlated during 2016/03-2016/04. For example, the correlation between \textbf{beta} and \textbf{residual volatility} was close to 1.0, which reflects that the factor returns were no longer independent and the model output was ineffective during the two months.}
\label{C3}
\end{figure}

\subsection{Model Effectiveness Measurement}

Investors who invest based on risk factors have to choose or develop a model to interpret the relationship between risk factors and returns. The models they choose are mostly statistical-based, which means the factors used in the model need to be updated frequently using the most recent data generated by the market. A strong correlation between factor returns may happen when the factors in the multi-factor model lose their effectiveness (\textbf{T1}). The Barra Risk model we adopt in this paper is no exception. The product managers($E_A$, $E_B$, $E_C$) from Ricequant realized that the data they feed into our system is calculated from a model that was created years ago (in 2012). They planned to use the system to find clues to verify the effectiveness of the factors used in the current version of the Barra Risk model (\textbf{T1}). 

To have an accurate comparison of the effectiveness of the factors between different time periods, product manager ($E_A$) slid a two-month time window in the comparison view to refresh the factor correlation view and observe the correlation between different factors (\textbf{T1}). The intensive red or blue color indicates a strong positive or negative correlation respectively. After spending some time looking over the factor correlation view corresponding to all the two-month time windows, he discovered that the colors in the factor correlation view were the most intense during March to April in 2016 compared to all other two-month time periods as shown in Fig.\ref{C3}. Several blocks in the factor correlation view showed intense red or blue color. He then suspected the factors that were generated from the Barra Risk model years ago were ineffective during the selected period. 

To prove his suspicion, he first slid the time window to the entire 2016 and then randomly selected several groups of portfolios from the portfolio cluster view to analyze them further in the comparison view (Fig.\ref{fig:teaser}$C$). All the groups selected were different in terms of risk strategy since their risk signatures were leaning towards different directions (\textbf{T4}). The top five industry holdings of the portfolios differed from each other (\textbf{T4}). $E_A$ clicked on several portfolio IDs to observe their stock holdings. It turns out some of the portfolios adopt a diversified strategy, holding many stocks on a low proportion, while others were much more concentrated. However, all the groups of portfolios had one thing in common. $E_A$ found that, in the aggregated return graph, almost all the portfolio return lines were fluctuating from March to April in 2016, indicating high volatility in the period. The returns calculated from the factor model should be independent in theory. In other words, portfolios holding different risk factors are supposed to enjoy the benefit of diversification. The return should be smooth and the volatility should be small. Therefore, the high volatility in the period proved the dependency between factors (\textbf{T1}). $E_A$ was then convinced that the model was indeed sometimes ineffective in the market.

\section{Discussion}

The case studies in the evaluation demonstrate the system's usability, and it has received positive feedback from the experts, confirming that we have succeeded in meeting the users' requirements. In addition, despite the fact that we used China A-Shares Market data in our visualization to prove the effectiveness of the system, it should be compatible with data from global markets as long as a risk factor model is adopted. In April 2019, we successfully conducted a pilot study where we put our system into production on RiceQuant's platform that serves over 100 Chinese financial institutions and 50,000 users. We plan to conduct further user surveys on a larger scale, as soon as our system launches, to glean further comments and information.

It is still possible to improve our system. Below we discuss three major limitations of the current system and the potential augmentations we might implement in order to apply the system to other generalized domains except for the financial field.

One limitation of our work is the scalability of the comparison view. In our case studies, users typically needed to compare around five clusters, each of which contained two to three portfolios. Our system can correctly handle such a demand. However, the current design would be significantly challenged, if a user was to observe an even larger number of portfolios simultaneously. This issue might arise, if our system were applied to other domains with a similar data format, or in another task in the financial field such as macro economy research.

Another limitation is that our system analyzes a portfolio in a stratified manner and then visualizes the stratum sequentially. Any insights concerning the inter-relationships between the strata are not visualized in our system. This is because the primary users of our system are quantitative traders who are more interested in the strata themselves rather than the relationships between them (this is background knowledge owned among traders). However, if researchers in the financial industry were to use our system, we would need to extend our designs to facilitate inter-stratum pattern exploration.

The final limitation is that whereas our system allows for the efficient exploration of stock portfolios, it is short of recommending how factors should be maintained or how a sector's influence on the portfolio should be maintained, in order to achieve the desired return. As of today, the industry professionals using our system will import the insights gained in our system into the other systems which will then generate stock recommendations. It would be more efficient for industry professionals and financial researchers if we were to merge these two functions together into a `one-stop-shop'.

\section{Conclusion}

In this work, we investigated the visual analysis of the stock portfolios. After eight months of close collaboration with a quantitative trading service provider company (RiceQuant) and two senior traders, three levels of analysis tasks were generalized, in order to better understand the stock portfolios, multi-factor models and the market situation. Our system was validated by three representative case studies, which served as a successful pilot study for commercializing our system with RiceQuant. A production schedule for our system is under intensive discussion, as well as a large user survey afterward. Our results indicated that \textit{sPortfolio} performs efficiently in stock portfolios exploration, and in trading strategy analysis, with multi-factor model interpretation. 

In the future, we plan to expand our system with stratum relationship analysis, which could benefit the automatic generation of quantitative investment strategies. Moreover, another portfolio except the stock one and other markets' data should be investigated to compare the performance of the same factor model, which could understand and empower the model comprehensively. In addition, we are considering implementing an automated recommendation system for portfolios. For example, predictions for favorable portfolios could be made via collaborative filtering\cite{herlocker1999algorithmic} based on the user ratings of each portfolio, and which could be measured by the similarity between user preferences (user-based) and portfolios (item-based). Indeed, other techniques that are commonly used in recommendation systems, such as neural collaborative filtering\cite{he2017neural} and deep learning models are all applicable in this context to empower our system.


\acknowledgments{
The authors thank the anonymous reviewers for their valuable comments and Dr. Rong Zhang for his continuous support on this work. This research work was partially supported by grant Hong Kong RGC GRF 16241916.

}

\bibliographystyle{abbrv-doi}

\bibliography{reference}
\end{document}